\documentclass[preprint2]{aastex}

\slugcomment{To appear in the ApJ, Oct 2006}
\def\hereafter{{hereafter~}}

\shorttitle{Blue Early-type Galaxies}

\shortauthors{Lee et al.}

\begin{document}

\title{THE NATURE OF BLUE EARLY-TYPE GALAXIES IN THE GOODS FIELDS}

\author{Joon Hyeop Lee, Myung Gyoon Lee, and Ho Seong Hwang}
\affil{Astronomy Department, School of Physics and Astronomy, FPRD, 
Seoul National University, Seoul 151-742, Korea}

\email{jhlee@astro.snu.ac.kr,
mglee@astrog.snu.ac.kr,
hshwang@astro.snu.ac.kr}

\begin{abstract}
We present a study of the nature of the blue early-type galaxies
(BEGs) in the GOODS north and south fields using the GOODS
HST/ACS archival data. Using visual inspection, we have selected
58 BEGs and 113 normal red early-type galaxies (REGs) in the
sample of 1,949 galaxies with spectroscopic redshifts. We find
that the BEGs are generally bluer, fainter, and less-massive than
the REGs, although a few BEGs are exceptionally bright and
massive. The number fraction of the BEGs to total early-type
galaxies is almost constant ($\sim0.3$) at $z \le 1.1$. In
addition, we find that the size
of the BEGs in given redshift bin decrease as redshift decreases.
The BEGs look similar to the REGs in the images and surface
brightness profiles. However, at least 27 BEGs show traces of
tidal disturbances in their fine structures: elongated cores,
off-centered cores, asymmetric internal color distributions,
tidally distorted outer structures, collisional rings, or very
nearby companions. Twenty-one BEGs are detected in the X-ray
bands and eleven of them are as luminous as $L_{0.5-10 {\rm keV}}
\ge 10^{43.5} ~{\rm erg~s}^{-1}$, indicating the existence of
AGNs in their centers. These results show that at least a half of
the BEGs may be descendants of mergers/interacting-galaxies and
that at least a quarter of the BEGs may be AGN-host galaxies. The
BEGs may evolve into REGs, and the size evolution of the BEGs is
consistent with the galactic \emph{downsizing} scenario.
\end{abstract}


\section{INTRODUCTION}

Early-type galaxies are one of the key objects in the modern
observational cosmology. One of the fundamental questions about
early-type galaxies is their formation history, which is closely
related to the history of cosmic mass assembly and cosmic
star-formation \citep{lee03}. There are two competing models for
the formation of massive elliptical galaxies. One is a monolithic
collapse model, according to which massive elliptical galaxies
were formed via a single collapse of proto-galactic cloud
followed by a rapid formation of stars \citep{par67,tin72,lar74}.
The other is a hierarchical merging model where massive galaxies
were formed through continuous merging and accretion of smaller
galaxies \citep{too77,sea78}.

The properties of early-type galaxies at low redshift are
relatively well known in many aspects thanks to a tremendous
number of studies for several decades: the surface brightness
profiles of early-type galaxies follow the $R^{1/4}$ law
\citep{dev48}; there is a strong correlation between luminosity
and central velocity dispersion of early-type galaxies
\citep{fab76b}; early-type galaxies show very slow rotation
\citep{ber75,ill77,dav83}; there are poor cold-gas contents in
early-type galaxies, implying rare star-formation events
\citep{fab76a}; the color and magnitude of early-type galaxies
are strongly correlated \citep{vis77}; there is a tight
correlation among size, surface brightness, and velocity
dispersion of early-type galaxies \citep[fundamental plane;
][]{luc91,jor93}; and the center of an early-type galaxy is
redder than its outer part \citep{vig88,fra89}. However, it is
still unclear whether all early-type galaxies have such `typical'
properties. Because understanding the property variation of
early-type galaxies along redshift makes it possible to constrain
the model for the evolution of early-type galaxies more strongly,
it is needed to observe and investigate early-type galaxies in
the deep universe.

Recent studies based on the HST images found a new class of
early-type galaxies in the distant universe that have different
properties from nearby early-type galaxies. \citet{abr99} showed
from the analysis of internal color dispersion that 40\% of
early-type galaxies (four out of eleven) at $0.4<z<1$ in the
Hubble Deep Field \citep[\hereafter HDF; ][]{wil96} have evidence
of current star-formation. \citet{men99} also identified three
star-forming E/S0 galaxies among the HDF galaxies. Subsequently,
\citet{men01a} estimated the surface color dispersion of the HDF
objects and reported that about 30$\%$ of 79 spheroidals show
strong internal-color variation, indicating the existence of not
only old stellar populations but also young stellar populations.
These objects generally show a bluer center than its outer part,
unlike typical nearby early-type galaxies. \citet{men01b} tried to
explain the origin of the blue centers with a multizone
single-collapse model, in which late infalls of material into the
high-potential core could cause prolonged star-formation in the
central region of a galaxy. Blue E/S0 galaxies were also found in
other HST fields. \citet{im01} found ten blue spheroids at
$0.2<z<1$ from the HST ``Groth Strip'' \citep{gro94} data. Based
on the small velocity dispersions of those spheroids,
\citet{im01} concluded that they are low-mass systems suffering
late star-formation events. Later, \citet{men04} identified
early-type systems with inhomologous internal colors (30\%--40\%
of 116 spheroids) in the HST/ACS UGC 10214 field using
photometric redshifts. They suggested that the internal color
distribution of these objects can be explained by recent
star-formations, and that the fraction of `active systems' does
not vary severely at $z < 1$. In addition, \citet{men05} pointed
out in a study of early-type galaxies in Abell 1689 that the
existence of an AGN in a galaxy is another possible origin of the
blue center in these galaxies.

Blue E/S0 galaxies were also found in two recent great missions:
UDF \citep[Hubble Ultra Deep Field; ][]{bec03} and GOODS
\citep[Great Observatories Origins Deep Survey; ][]{dic03}.
\citet{elm05a} found 30 early-type galaxies with blue central
clumps in the UDF, and estimated the sizes, masses, and ages of
six galaxies with spectroscopic redshifts. They found that the six
galaxies are 1 -- 5 Gyr old (and their clumps are younger than
100 Myr) and a half of them are more massive than $10^{10}$
M$_{\odot}$, although there could be some degeneracy and
model-dependence in those estimations. \citet{fer05} analyzed the
statistical properties of 249 early-type galaxies in the southern
GOODS field, and reported that one third of the early-type
galaxies at intermediate redshifts have blue colors. Based on the
analysis of the color gradients, color scatters, and
redshift-variations, they concluded that there exist a clear cut
between blue E/S0 galaxies and red E/S0 galaxies in their color,
and that the `two-burst formation scenario' can reproduce the blue
E/S0 galaxies. However, fine structures of these blue galaxies
were rarely investigated in previous studies.

In this paper, we present a study of the nature of the blue
early-type galaxies in the GOODS north and south fields, based on
the GOODS HST/ACS archival data.
We use only galaxies with spectroscopic redshift and focus on studying
the fine structures of the target galaxies.
The outline of this paper is as
follows. Section 2 describes the data set we used, and \S3
explains  our selection process for blue early-type galaxies. We
present the results of our photometric, structural, and
statistical analyses of the blue early-type galaxies in \S4.
Primary results are discussed in \S5, and a summary and
conclusions are given in \S6. Throughout this paper, we adopted
the cosmological parameters: $h=0.7$, $\Omega_{\Lambda}=0.7$, and
$\Omega_{M}=0.3$.

\section{THE DATA}

The GOODS is a survey project to observe deep and wide universe
with the most powerful facilities covering a large range of
wavelength. As a part of the GOODS program, the HST Treasury
program was carried out to obtain deep optical images of two
fields (the Hubble Deep Field North (HDFN) and the Chandra Deep
South (CDFS)) with four filters: F435W($B$), F606W($V$),
F775W($i$), and F850LP($z$). The total exposure times are 7,200,
5,000, 5,000, and 10,660 seconds for $B,V,i$, and $z$,
respectively, and the total spatial coverage is about 300
square-arcminutes. The pixel scale of the released
drizzled-images is $0.03$ arcsec per pixel, and the average PSF
FWHM is about $0.1$ arcsec. We used the GOODS HST/ACS images and
photometry catalogs (version 1.1) publicly released on 2004 April
9 \citep{gia04}. Among the objects in the photometric catalogs,
we selected galaxies whose spectroscopic redshifts are available
in the literature
\citep{coh00,cri00,cro01,daw01,bun03,ste03,cow04,dic04,lef04,sta04a,sta04b,str04,szo04,van04,van05}\footnote{The
master compilation of GOODS/CDFS spectroscopy is available at
http://www.eso.org/science/goods/spectroscopy/CDFS\_Mastercat/}:
1,017 galaxies in the HDFN and 932 galaxies in the CDFS (1,949
galaxies in total). For the CDFS galaxies, their spectra in the
\emph{GIF} format were retrieved from the
web-site\footnote{http://archive.eso.org/wdb/wdb/vo/goods\_CDFS\_master/form}
provided by \emph{ESO Astrophysical Virtual Observatory}, and
were used for inspecting spectral line features. In addition, we
used the X-ray data in two \emph{Chandra} observational catalogs
\citep{gia02,ale03} for the selected galaxies. We matched optical
objects with X-ray sources within 0.5 arcsec positional tolerance
that is almost the same as the pixel scale of \emph{Chandra}
images.

\section{SAMPLE SELECTION}

\subsection{Selection of Early-type Galaxies}

From the data set of 1,949 galaxies with spectroscopic redshifts,
we selected early-type galaxies by visual inspection of the
images. Because these GOODS galaxies are at various redshifts
from $z \approx 0$ to $z \approx 5$, the rest-frame wavelength of
one galaxy at given redshift can be largely different from that of
another galaxy at different redshift in the same band. To avoid
the selection effect that can be caused by the morphology
dependence on wavelength, we inspected all four band images of
each galaxy considering its rest-frame wavelength, according to
the following steps.

First, we set five redshift bins: 0 -- 0.087, 0.087 -- 0.515,
0.515 -- 0.938, 0.938 -- 1.125, and beyond 1.125. Each redshift
bin is set so that a range in which the rest-frame central
wavelength of each band ($B$, $V$, $i$, and $z$) is not shorter
than 4,000 {\AA} break at which a sudden spectral break occurs
due to old stellar populations. For example, the rest-frame
central wavelength of the $B_{435}$ band at $z=0.087$ is exactly
4,000 {\AA} and the rest-frame central wavelength of the
$V_{606}$ band at $z=0.515$ is also 4,000 \AA. Therefore, in the
redshift range from 0.087 to 0.515, the $V$, $i$, and $z$ bands
show rest-frame optical images but the $B$ band shows images of
$\lambda_{0} < 4,000$ {\AA}. After setting the redshift bins, we
conducted visual classification using all four band images for
each object. We considered as early-type galaxies those which
look circular or elliptical, but with little hint of disk in the
bands of $\lambda_{0} \ge 4,000$ {\AA}. Since rest-frame optical
images at $z>1.125$ in any GOODS band were not available, we
classified as early-type those galaxies that have circular or
elliptical morphology in the $z$ band.

We repeated this procedure three times and determined the
final classes with the median of the three independent
classification results.
The number of galaxies finally classified as early-type is 171.
We compared our classification to that of \citet{bun05} who classified
galaxies visually in the same fields. It is found that $88\%$ of our
early-type galaxies were classified as E, E/S0, compact galaxy, or star
by \citet{bun05}, but only $62\%$ of the objects that ranked as E, E/S0,
compact galaxy, or star by \citet{bun05} are classified as early-type in
our study. Three objects in our sample of early-type galaxies classified
as stars by \citet{bun05}  are not stars 
but probably compact galaxies, because they have spectroscopic redshifts larger than $0.3$.
This comparison shows that the classifications in the two
studies are consistent, but that our sample was selected more strictly.
The small disagreement is partially due to the different criterion
for the UV images at the rest-frame wavelength.

Because we selected early-type galaxies in a sample combined from 
various redshift surveys, our sample is heterogeneous.
The number of redshift surveys referred here is fifteen, and a major
fraction of redshift data at $z<2$ were retrieved
from five surveys \citep{coh00,cow04,lef04,szo04,van05}.
These five surveys have slightly different flux limits in selecting
spectroscopic targets: $R=24$ in \citet{coh00}, $R=24.5$ in \citet{cow04},
$I=24$ in \citet{lef04}, $R=26$ in \citet{szo04}, and $z=24.5$ in
\citet{van05}. In four redshift surveys except \citet{van05}, targets
were selected with nearly unbiased methods (basically magnitude limited).
In the sample of \citet{van05}, main targets were selected
among the galaxies with $(i-z)>0.45$, which may be somewhat biased to
red galaxies (possibly early-type galaxies).
Nevertheless, our sample is still useful in probing the nature of
blue early-type galaxies.

We discuss a possible bias in using only-spectroscopic sample.
Even if a target selection itself in a spectroscopic survey is not biased,
it is plausible that the success rate in securing spectra is higher
in objects with strong lines than in objects with weak lines.
This can result in the excess of the late-type ratio in the whole sample, or
in the case of our study, the excess of the blue-early-type ratio in the early-type
sample (if we suppose that the blue early-type galaxies have typically stronger
lines than the red early-type galaxies). This is an unavoidable effect as long
as we use spectroscopic sample. 
Photometric redshifts could be used as well, but current estimates of
the photometric redshifts of the galaxies in the GOODS fields are not perfect.
In particular it is not easy to derive a reliable estimate of the redshift for blue
early-type galaxies that are main targets in this study. 
We decided to use only the spectroscopic redshifts because we needed accurate
values of the redshifts for the target galaxies.

The fraction of the early-type galaxies in our sample is $9\%$ (=171/1949).
This is somewhat smaller than the result of \citet{bun05} 
who classified the galaxies visually
using the same images of the GOODS field, obtaining $14\%$ (E, E/S0 in their sample).  
Our result is similar to the results of 
\citet{elm05b} who classified the galaxies visually
using the HST/ACS images of the UDF, obtaining $11\%$.
However, it is smaller than those based on the automatic classification:
\citet{con05} obtained $20-30\%$ at $z\sim0.5$ in the GOODS fields.
In the cases of \citet{con05}, however, early-type 
galaxies were selected using a quantitative methods (concentration-asymmetry-clumpiness
correlations), which possibly regard bulge-dominant late-type galaxies as
early-type.

\subsection{Selection of Blue Early-type Galaxies}

We divided the sample of early-type galaxies into two groups: red
early-type galaxies (REGs) and blue early-type galaxies (BEGs),
using the $(i-z)_{\rm AB}$ color variation as a function of
redshift. REGs correspond to typical early-type galaxies seen in
the local universe. The color of early-type galaxies varies
significantly as redshift increases but the color evolution of
typical REGs are approximately reproduced with the simple stellar
population (SSP) model. Therefore, it is efficient to use the
color difference between the observed color and the SSP
expectation value for selecting BEGs.

Fig.~1 displays the $(i-z)$ color difference ($\Delta (i-z)
\equiv (i-z)_{\rm model} - (i-z)_{\rm obs}$) vs. redshift for the sample of
early-type galaxies. To derive $(i-z)_{\rm model}$, we used the
SSP model with formation redshift $z_{\rm F}= 5$ (single-burst
12.4 Gyr ago) and metallicity [Fe/H] $= -0.4$ dex, calculated
with the GALAXEV code \citep{bru03}. [Fe/H] $= -0.4$ dex was
chosen because it matches well the observed colors of REGs.
Several features are noted in Fig.~1. First, most galaxies at
$z<1.2$ are concentrated around the zero color difference,
showing that their colors are well reproduced by the SSP model.
Second, there are a significant number of galaxies with large
positive color differences at $z<1.2$, and all galaxies at
$z>1.2$ show even larger color differences. Third, one galaxy at
$z\approx 0.75$ shows a large negative color difference,
indicating that it may be a highly reddened galaxy.

In Fig. 1, we find no early-type galaxy that agrees with the SSP
expectation at $z>1.2$. This is because only rest-frame UV images
can be seen at $z>1.2$ in the GOODS bands and because typical
early-type galaxies (i.e., REGs) are very faint in the UV
wavelength.

Considering the features seen in Fig. 1, we divided the sample of
galaxies at $z\le1.2$ as follows. First, we divided the redshift
range of $0 \le z \le 1.2$ into five bins, and derived a color
histogram of the galaxies for each redshift bin. In the left
panel of Fig.~2, the redshift and $\Delta (i-z)$ values of the
galaxies at $0\le z \le 1.2$ are plotted. We divided this
redshift range into five bins so that each bin contains a similar
number of galaxies: 0 -- 0.4, 0.4 -- 0.5, 0.5 -- 0.6, 0.6 -- 0.8,
and 0.8 -- 1.2. For each bin, we present an $(i-z)$ color
difference histogram in the right panel of Fig.~2. The color
difference histogram in each bin shows a strong peak around the
$\Delta (i-z)=0$, which can be fit well by a gaussian function.
The galaxies around this peak correspond to the REGs. We derived
a guideline for color separation by fitting a 3rd-order
polynomial function to the colors corresponding to the peaks in
the five bins. We regard as BEGs the early-type galaxies that are
$3 \sigma$ bluer than the peak color. Applying the criteria
established above, we selected 113 REGs and 58 BEGs.
This selection is not sensitive to the choice of evolutionary models,
because the main criterion is the
color distribution itself of early-type galaxies.

We checked the possibility that our BEG sample is biased to S/Irr
types of \citet{bun05} sample, and found that this is not
that case: $64\%$ of our BEGs are classified as E, E/S0, compact galaxy,
or star by \citet{bun05}, which is just a little smaller than
the ratio for our entire early-type sample.

Fig.~3 displays the $(i-z)$ color vs. redshift diagram of the
REGs and the BEGs. We also plotted 1,778 galaxies classified as
non-early-type for comparison. We plotted as well several
evolutionary models with [Fe/H] $= -0.4$ and $z_{\rm F} = 5$
given by \citet{bru03}: an SSP model and three
exponentially-decreasing star-formation rate models with
exponential time-scales, $\tau = 0.5, 1,$ and 8 Gyr.

Fig.~3 shows several important features of the REGs and the BEGs.
1) The REGs follow closely the SSP model up to $z\approx 1.2$. 2)
The BEGs are found at $0<z < 3.7$, while the REGs are found only
at $z< 1.2$. 3) The color dispersion of the BEGs at given redshift
is larger (ranging from $(i-z)\approx 0.2$ to 0.75 at $z \approx
1$) than that of the REGs, as also seen in Fig.~2. 4) All BEGs at
$z>1.2$ are significantly bluer than the SSP model and their
color evolution matches roughly the exponentially-decreasing
star-formation rate models rather than the SSP model.

Table~1 lists the basic properties of the BEGs selected in this
study: our short ID, IAU ID, spectroscopic redshift, and AB
magnitudes in four bands. We used short IDs of `GN' for GOODS-HDFN
and `GS' for GOODS-CDFS. The maximum photometric errors in our
BEG sample are 0.055, 0.027, 0.027, and 0.015 for $B,V,i$, and
$z$, respectively.

\section{RESULTS}

We have derived several physical parameters of the 58 BEGs and
have investigated their morphological structures in comparison
with those of the REGs.

\subsection{Color-Magnitude Diagram}

Fig.~4 displays the $(i-z)_{\rm AB}$ vs. $z_{\rm AB}$ diagram of
the BEGs in comparison with the REGs and non-early-type galaxies
selected in this study. Fig.~4 shows following features. 1) The
magnitudes of most BEGs range from $z_{\rm AB}\approx$ 19.5 to
23.7, and two BEGs (GN 14242 and GS 895) are about 1.5 mag
brighter than the rest of the BEGs. 2) The magnitudes of the REGs
are in the similar range to those of the BEGs. 3) However, the
BEGs are relatively bluer and fainter than the REGs so they can be
roughly separated even in the observed color-magnitude diagram.
4) The colors of the REGs get redder as their redshift increases,
while those of the BEGs at the low redshift and high redshift
occupy a similar range.

\subsection{Magnitudes and Sizes vs. Redshift}

We have investigated the variation of the absolute magnitudes and
sizes of the BEGs as a function of redshift in comparison with
the REGs, which are listed in Table 2. Fig.~5 displays apparent
$z$-band magnitude ($m_{z,obs}$), absolute $z$-band magnitude
($M_{z, obs}$), half-light radius (R$_{hl,z}$) in pixel and
R$_{hl,z}$ in kpc. The typical estimation error of half-light
radius, err($R_{hl}$) is about $+4\%$/$-5\%$.
We present the mean magnitudes and sizes of
the BEGs and REGs as a function of the redshift in Fig.~5. Here
we used the observer-frame magnitudes instead of the rest-frame
magnitudes, because the spectral energy distributions (SEDs) of
the BEGs are too diverse to derive K-correction.

Several interesting features are noted in Fig.~5. 1) The absolute
$z$-band magnitudes of the BEGs are fainter on average than those
of the REGs. The differences in the maximum absolute $z$-band
magnitudes between the BEGs and the REGs get smaller as the
redshift inceases, reaching almost zero at $z\approx 0.9$. The
lower envelopes of the absolute magnitudes of the BEGs are
affected by the magnitude limit, but the upper envelopes should
be free of this. 2) The mean values of the half-light radii of
the BEGs are smaller than those of the REGs. 3) While the
half-light radii of REGs show little change as redshift increases
until $z\approx 1.2$, those of the BEGs increase along redshift.
4) The half-light radii of the BEGs at $z>1.2$ are smaller than 2
kpc. 5) Two brightest BEGs at $z\approx 0.5$ and $\approx 0.85$
(GN 14242 and GS 895, respectively) are brighter than the
brightest REGs at the same redshift, and their half-light radii
are as large as the largest REGs.

Fig.~6 displays the absolute magnitude vs. the half-light radius
of the REGs and the BEGs. It shows that both REGs and BEGs get
brighter as they get larger. One remarkable feature in Fig.~6 is
that the BEGs at $z>1.2$ do not follow this correlation. They are
very luminous in spite of their small sizes, showing a possibility
that most BEGs found at $z>1.2$ may contain AGNs.

We estimated roughly the masses of the sample galaxies, using
galaxy evolution models of \citet{bru03} overlaid on the
magnitude vs. redshift diagram, as shown in Fig. 7. Three kinds
of galaxy evolution models were used: one SSP model with $z_{\rm
F} = 5$, one SSP model with $z_{\rm F}= 2$, and one
exponentially-decreasing star-formation rate model with $z_{\rm
F}= 5$ and $\tau=8$ Gyr. In all models, the metallicity [Fe/H]
=$-0.4$ was adopted. From the comparison of the data with the
evolution models, we find the followings. 1) Using the SSP model
with $z_{\rm F}=5$, we estimate the mass range of the REGs to be
from $10^{10} {\rm M}_{\odot}$ to $10^{12} {\rm M}_{\odot}$. Most
REGs have masses from $10^{11} {\rm M}_{\odot}$ to $10^{12} {\rm
M}_{\odot}$. 2) It is not easy to estimate reliably the masses of
the BEGs, because their star formation histories are not yet
known. We used one SSP model with $z_{\rm F}=2$ and one
exponentially-decreasing star-formation rate model with $z_{\rm
F}= 5$ and $\tau=8$ Gyr for this. Using the SSP model with
$z_{\rm F}=2$, we find the masses of most BEGs range from $10^{10}
{\rm M}_{\odot}$ to $10^{12} {\rm M}_{\odot}$. More than half of
them have masses lower than $10^{11} {\rm M}_{\odot}$, and some
even lower than $10^{10} {\rm M}_{\odot}$. In summary, the BEGs
have relatively lower masses than the REGs.
However, it should be considered that faint spectra without 
strong lines (e.g. faint REGs) are harder to secure than faint 
spectra with strong lines (possibly faint BEGs), and that
the superiority in magnitude, size and mass of REGs to BEGs shown
in Fig. 5 and Fig. 7, can be partially caused by this spectroscopic bias. 

Fig.~8 displays the number counts of the REGs and the BEGs, and
the variation of the BEG fraction in the entire sample of
early-type galaxies along redshift. The redshift distribution of
the BEGs is almost flat at $z\le1.2$, whereas that of the REGs
show a strong concentration at $z \approx 0.55 $. 
\citep[this concentration may be originated from the large-scale
clustering seen in the GOODS fields;][]{con05}. No REG was
found at $z>1.2$ due to the limitation of optical observation,
as mentioned in \S3.2. The fraction of the BEGs to entire
early-type galaxies is almost constant ($\sim 0.3$) from $z=0.1$ to $z=1.1$, which is in  good agreement
with the result of \citet{men04} based on the study of HST/ACS
images of the UGC 10214 field. Noting that the target fields and
the sample selection are different in these two studies, 
the agreement between the two results is remarkable.
This indicates that the fraction of the BEGs in early-type galaxies
may be constant at $z<1.1$.

In Fig.~9, the half-light radius distribution (left panel) and
the absolute magnitude distribution (right panel) of the REGs and
the BEGs, respectively, are presented for four redshift bins. At
first glance, these histograms show that the BEGs at lower
redshift are smaller and fainter than the BEGs at higher
redshift, unlike the REGs that keep almost consistent size and
luminosity regardless of their redshift. However, we need to
consider the following two points. First, since we did not
conduct k-correction (neither photometrically nor
morphologically), it can be affected by selection effect to
compare sizes or magnitudes between different redshift bins
directly. Therefore, we should focus on the \emph{difference}
between the BEGs and the REGs in each redshift bin, not the
values themselves. Second, because magnitudes are dependent on
the star-formation history as shown in Fig.~7, the magnitudes of
REGs and BEGs may be different systematically if the BEGs have more
recent-star-formations than the REGs. Moreover, exact k-corrections
are necessary for fair comparison of magnitudes, which is
hard to conduct for the BEGs. On the other hand,
the size of a galaxy is affected by its dynamical history but
relatively less affected by its star-formation history.
When supposing typical velocity dispersions, size may represent 
the mass of a dynamically-stable system straightforwardly.

Considering those points, we compared the median sizes of the BEGs
and the REGs in four redshift bins in Fig. 10. The upper-left
panel of Fig.~10 displays how the median sizes of the REGs and
the BEGs vary along redshift. In the $z$ band, both the REGs and
the BEGs enlarge as redshift increases. The size of the BEGs in
the $B$ band also shows an increasing trend along redshift,
although there is a small reversion from $z \sim 0.5$ to $z \sim
0.7$. However, the size of the REGs in the $B$ band becomes
smaller as redshift increases. This is because the $B$ band
represents rest-frame UV wavelength at $z\sim 1$ and because
typical REGs are very faint in the UV wavelength due to the lack
of young stellar population.

In the lower-left panel of Fig.~10, the ratio of BEG median size
to REG median size as a function of redshift is displayed. 1) In
the $z$ band, the BEG/REG median-size ratio evolves from 0.52 at
$z \sim 1.0 $ to 0.36 at $z \sim 0.2 $. 2) In the $B$ band images,
the size ratio evolves more rapidly than in the $z$ band,
resulting that the size ratio at $z \sim 0.2$ is only 0.28 in
spite of the extremely high ratio at $z \sim 1.0$. Considering
the rest-frame wavelength of each redshift bin in each band, the
$B$-band size ratio at $z \sim 0.2$ (0.28) that is even smaller
than the $z$-band size ratio at $z \sim 1.0$ (0.52), confirms the
size -- redshift relation of the BEGs, free from the dependence
of galaxy size on wavelength.  

In addition to the wavelength effect,
there is one more thing to be considered: the lower limit of 
observable size. If there are more faint BEGs undetected at 
$z\sim 1$ than faint REGs (although it is an opposite supposition
to the possible spectroscopic bias mentioned above), 
this `small BEG missing' can cause
the overestimation of the BEG median size at $z \sim 1$. 
To avoid this effect, we also used the 3rd largest galaxy sizes for both
the REGs and the BEGs, which represent the upper envelope
of galaxy size in each bin (the largest or the 2nd largest can be
very sensitive to the existence of abnormally large objects).
Since the 3rd largest galaxies are much brighter than the flux limit of
any survey we referred up to $z\sim1$ (by $3-4$ mag; see Fig. 5), 
this parameter is almost free from observational biases.
The result is presented in the upper-right
and the lower-right panels of Fig.~10, which shows that the size
ratio varies more rapidly than the median values. This result 
shows that the size evolution of the BEGs is not a selection
effect caused by the lower envelope of observable magnitude. In
summary, the BEGs at low redshift are \emph{indeed} smaller (and
maybe less massive) than the BEGs at intermediate redshift
($z\sim 1$).
Recently, \citet{ilb06} showed a similar trend,
based on the evolution of galaxy luminosity function 
according to morphology up to $z=1.2$ using VIMOS-VLT Deep Survey
\citep{lef04} and COMBO-17 \citep{mei02} data: 
`blue bulge-galaxies' are dimming by 0.7 mag from $z=1$ to $z=0.6$, 
which is in a good agreement with the size evolution of BEGs 
in our sample.

\subsection{Structure}

We display the gray-scale maps and isophotal contours
of $z$-band images of the 58 BEGs in Fig.~11. Three typical REGs (GN10606,
GN15469, and GN22886) with different apparent magnitudes ($m_{z}
= 20.85$, 18.87, and 22.01 respectively) are also displayed for
comparison.

Fig.~11 shows that most BEGs look in general very similar to the
REGs so it is difficult to distinguish them without colors. Only
a few BEGs show slightly distorted or elongated isophotes in the
central regions (GN14128, GN17606, GN19857, GN25594). We fitted
the $z$-band images of the BEGs using the IRAF/ELLIPSE task.
Fig.~12 displays the surface brightness profiles of the BEGs and
three REGs for comparison. Since the profiles within the PSF
radius are flat, we fitted the surface brightness profiles of
outer regions of the galaxies with the de Vaucouleur ($R^{1/4}$)
law. In Fig.~12, the surface brightness profiles of outer regions
of most BEGs are fit well by the de Vaucouleur law like those of
the REGs. Thus there is little difference in the images and
surface brightness profiles between the BEGs and the REGs.

\subsection{Color Maps}

We created pseudo-color maps representing the $(V-i)$ color of the
BEGs using $V$ and $i$ images, as shown in Fig.~13.
Strikingly diverse features of the BEGs are noted in Fig. 13.

A blue clump (BC) mostly in the central regions of the galaxies
is one conspicuous feature of many BEGs. There
are several types of blue clumps found in Fig.~13: a symmetric
deep BC (i.e., a steep blue dip in the color profile; e.g.,
GN1430, GN5302, and GS20457); a symmetric shallow BC (i.e.,
shelving inverse color gradient in the color profile; e.g.
GN14229, GS13482, and GS21636); an asymmetric (or elongated) BC
(e.g., GN8292, GN21983, and GS9479); an off-centered BC (e.g.
GN10598, GN16895, and GS7517); and multiple BCs (GN28407). For a
typical massive REG, it is well known that its color profile
shows a negative color gradient (i.e., its central part is redder
than its outer part) probably originated from its internal
metallicity gradient \citep{tam00}, and a blue clump is hardly
found in bright REGs. A blue clump may represent the existence of
a large amount of young stellar population or, for some deep BCs,
may represent the existence of an AGN in the center of the BEG.

In addition to blue clumps, other kinds of remarkable features are
found in the outer regions of BEGs. Although the intensity maps of
BEGs show good elliptical or circular shapes in Fig.~11, the
pseudo-color maps in Fig.~13 reveal diverse faint structures in
the BEGs: 1) tidally distorted outer structures (e.g., GN1430,
GN9126, and GN10598); 2) very nearby (and maybe interacting)
objects (e.g., GN5302, GN7904, and GS17085); 3) a red inner ring
(e.g., GN19996, GS10754, and GS20457); and 4) a blue outer ring
(GN11421). The tidally distorted outer structures and the very
nearby companions indicate that those BEGs may be under the tidal
influence from other objects, probably responsible for the blue
stellar population.

The nature and origin of the red inner ring are difficult to explain. 
One possibility is that they are artificial features resulted 
from the PSF variation according to ACS bands (see Appendix B in \citet{men04}).
However, the red inner ring features are conspicuous both in the $(V-i)$ color
maps and in the $(V-z)$ color maps, and their positions and sizes are
almost identical in the two kinds of color maps. If they are indeed
the results of PSF variation, the shapes of the rings in each color map
may be significantly different from each other. Therefore, the red 
inner rings are considered to be real features.

GN11421, the BEG with a blue outer ring, shows an interesting
morphological variation along wavelength. The properties of the
ring galaxies at high redshift including not only GN11421 but also
two other GOODS objects will be described in the Appendix. A
possible origin of the ring feature of GN11421 is a head-on
collision between two galaxies \citep{app96}.

We used the $(V-i)$ color maps to investigate the fine color 
structures of BEGs, to minimize the
effects caused by different PSF FWHMs in different bands. It is
generally known that the ACS $z$-band FWHM is wider than those of
$B,V,$ and $i$ bands \citep{men04}. 
(The reason we did not use the $B$ band
images is that many BEGs are too faint in the $B$ band.)
However, since $(V-z)$ color can show fine color structures more
conspicuously than $(V-i)$ color owing to its wide wavelength
interval, we checked the difference between the features in the 
$(V-i)$ color maps and those in the $(V-z)$ color maps. Most major
features are not much different in the two color maps, but
the shapes of some central BCs look slightly different in different
colors. Some asymmetric features of central BCs are too
conspicuous to be artifacts, but some may result from 
the effect of PSF variation according to bands.
In the $(V-z)$ color map, it is interesting that GN14128 with 
an off-centered core in the contour map has a blue clump biased 
toward the opposite direction to the bright core seen in the $z$ 
band, which is good evidence for the tidal event of 
GN14128 and is not seen in the $(V-i)$ color map.

Fig.~14 displays the $(V-z)$ color profiles of the 58 BEGs and
three REGs. We inspect the $(V-z)$ color profiles not the $(V-i)$ 
color profiles, because it is more advantageous to use colors with
a wider wavelength interval in investigating radial variations of 
stellar population, and because the PSF variation effect may be not 
significant except within the small central regions in the surface
profiles of galaxies.
In Fig.~14, the brightest one (GN15469; $M_{z} = -22.13$) among
the three REGs shows almost flat color profile and the
intermediately-bright REG (GN10606; $M_{z}=-21.33$) also has an
almost flat color profile with small scatters, while the faintest
one (GN22886; $M_{z}=-18.75$) shows a very weakly-blue center.
Compared with these REGs, the BEGs have very diverse color
profiles: some BEGs (e.g., GN14242, GN16895, and GS895) are not
much different from the bright REGs; some BEGs (e.g., GN9126,
GN14128, and GS13482) have just shallow BCs like the faint REG;
and some BEGs (e.g., GN5302, GN19996, and GS20048) have very deep
BCs that are not found in the REGs. In addition, the color
profiles of some BEGs do not vary monotonically: the color
profiles in some BEGs vary from increasing to constant (e.g.
GN13334 and GS20048); the color profiles in some other BEGs vary
from constant to increasing (e.g., GN17217 and GS28929); and the
color profiles in other BEGs undulate (e.g., GS6587 and GS11796).
The complex color profile may indicate the complex spatial
distribution of the stellar populations in the BEG. The features
found in the pseudo-color maps are summarized in Table~3.

\subsection{X-ray Luminosity and Spectral Line Feature}

We calculated the X-ray luminosity of the BEGs that have X-ray
flux data, adopting a typical AGN SED, as listed in Table 3. It
is generally known that an AGN generates strong X-ray emission
\citep{elv78} and that vigorous star-formation activities can
also lead to X-ray emission \citep{lai05}. The criterion is somewhat
ambiguous, for distinguishing between X-ray emission from an AGN
and X-ray emission from star formation activity. We divided the
sample of BEGs into low X-ray luminosity objects (L$_{0.5-10 {\rm
keV}} <10^{43.5}$ erg ${\rm s}^{-1}$) and high X-ray luminosity
objects (L$_{0.5-10 {\rm keV}} \ge 10^{43.5}$ erg ${\rm
s}^{-1}$). This separation is based on the fact that the maximum
X-ray luminosity of the `spectroscopic non-AGNs' (see below) in
our sample is $10^{43.34}$ erg ${\rm s}^{-1}$. Fig. 15 shows
these X-ray luminosity features on the $\Delta(i-z)$ vs. redshift
diagram. The X-ray BEGs are found only at $z>0.5$ and their
$\Delta(i-z)$ values range from very small ($\approx 0.15$) to
very large ($\approx 1.15$). It is remarkable that nine of the
thirteen BEGs at $z>1.2$ have high X-ray luminosity, implying
that most BEGs at $z>1.2$ may be probable AGN-host galaxies.

Using the spectra, we classified the CDFS BEGs according to their
spectral line features into four categories: absorption line
galaxies, emission line galaxies, composite line galaxies, and
AGNs, as listed in Table 3. Fig.~16 displays the spectral line
features on the $\Delta (i-z)$ vs. redshift diagram. The following
features are noted. 1) There are six absorption-line BEGs: four
with $(i-z)<0.35$ at $z<0.8$ and two with $0.5<(i-z)<0.6$ at
$z\approx1$ and $1.5$. 2) Non-AGN objects are found only at
$\Delta(i-z)<0.6$. 3) Some emission-line BEGs and composite-line
BEGs have smaller $\Delta(i-z)$ than the average $\Delta(i-z)$ of
absorption-line BEGs. Since emission lines in a galaxy are
generally a good indicator of the star-formation activity or AGN
activity, this indicates that most BEGs probably contain ongoing
star-formation or an AGN.

We checked the correlation between the blue clumps and the X-ray
luminosity, finding a significant coincidence between `a symmetric
deep blue clump or an elongated blue clump' and `X-ray emission'.
There are eleven BEGs with L$_{0.5-10 {\rm keV}} \ge 10^{43.5}$
erg ${\rm s}^{-1}$ (hereafter, \emph{X-ray luminous}) and 16 BEGs
with a deep blue clump (hereafter, \emph{blue clumpy}). 
Among them, ten BEGs are both \emph{X-ray luminous} and 
\emph{blue clumpy}; in other words, $91\%$ of \emph{X-ray luminous}
BEGs are \emph{blue clumpy} and $63\%$ of \emph{blue clumpy} BEGs
are \emph{X-ray luminous}.

We also checked the correlation between the X-ray luminosity and
the spectral line features, finding four of eight `spectroscopic
AGNs' (GS467, GS3739, GS20244, and GS22653) are \emph{X-ray
luminous} BEGs. One spectroscopic AGN GS20457 did not match any
X-ray source, and the other three spectroscopic AGNs (GS6961,
GS13482, and GS20048) are in the X-ray luminosity range of
$10^{42.35}\le$ L$_{0.5-10 {\rm keV}} \le 10^{43.14}$, implying
possible existence of more AGNs with low X-ray luminosity.

\section{DISCUSSION}

\subsection{The Nature of the BEGs}

In the previous studies, blue E/S0 galaxies were regarded as: 1)
star-forming low-mass spheroids with inverse color gradients
\citep{im01}; 2) early-type galaxies with a blue central clump
caused by prolonged single-collapse \citep{men01b} or caused by
the accretion of blue objects \citep{elm05a}; 3) early-type
galaxies that suffered the secondary starburst \citep{fer05}; or
4) AGN-host galaxies \citep{men05}. In this study, we found the
followings, focusing on the fine structures of the blue early-type
galaxies.

First, at least $47\%$ ($27/58$) of the BEGs have traces of tidal
interactions (an elongated or off-centered core; an elongated or
off-centered blue clump; multiple blue clumps; tidally distorted
outer structures; interacting companions; or a blue outer ring).

Second, at least $28\%$ ($16/58$) of the BEGs probably contain
AGNs in their centers (high X-ray luminosity or AGN spectra).
This result supports the idea that AGNs may be responsible for the
blue color of some blue E/S0 galaxies \citep{men05}. It is
interesting that $14\%$ ($8/58$) of the BEGs (GN1430, GN5302,
GN8292, GN24520, GN24667, GS3739, GS20048, and GS22653) show both
the traces of tidal events and the AGN features, because this
partially supports the merger origin of AGNs
\citep{can01,kew03,san05}.

Third, $19\%$ ($11/58$) of the BEGs have red inner rings whose
nature is still unclear. All of these red-ring BEGs are
\emph{blue clumpy} and $69\%$ ($11/16$) of the \emph{blue clumpy}
BEGs have red inner rings. This indicates that the red inner rings
are possibly related to AGN activity or strong star-formation
activity. However, it is not easy to explain how such rings form
in detail.

Fourth, as described in \S4.2, the BEGs have smaller sizes than
the REGs on the average. However, there are also several BEGs
with large sizes: $22\%$ ($13/58$) of the BEGs are larger than
$R_{\rm hl,z}=2$ kpc and two BEGs (GN14242 and GS895) are even as
large as $R_{\rm hl,z}\approx 10$ kpc. Considering such large
sizes of those BEGs, they are hardly `star-forming low-mass
spheroids \citep{im01}' unless they are very unstable dynamically
or their velocity dispersions are unusually small.

Almost a half of the BEGs show evidence of tidal events and at
least a quarter of the BEGs probably contain an AGN in their
center. At least in our sample, the role of extra-galactic
disturbance appears to be more dominant than that of a prolonged
single collapse \citep{men01b}. Within the frame of the
`two-burst model' \citep{fer05}, the `secondary burst' may be
interpreted as the result of a tidal interaction or merging with
other objects.

\subsection{BEGs and Galactic \emph{Downsizing} model}

Recently, evidences are accumulating, supporting a new scheme for
early-type galaxy formation: the galactic \emph{downsizing}
\citep{cow96,sus00,tre05}. According to the \emph{downsizing}
model, the more massive galaxies are, the older they are. Although
this scenario started as a modified version of the monolithic
collapse model, it was suggested more recently that the
\emph{downsizing} can be also compatible with the hierarchical
merging scenario. \citet{bun05} found that the E/S0 fraction
decreases as redshift increases, while the stellar mass function
for all galaxies shows little evolution. At the same time, they
also found that the lower-mass limit of E/S0 galaxies is higher
at earlier epoch, which indicates that galaxy merging may be
significantly related to the \emph{downsizing} scheme.
\citet{fab06} supported this idea with the evolution of the
luminosity function for blue and red galaxies to $z\sim1$ using
the DEEP2 \citep{dav03} and the COMBO-17 \citep{wol03} survey
data. According to them, the `quenching mass' from blue galaxies
to red galaxies via gas-rich mergers, decreases as time goes.
\citet{del06} showed that it is possible to reproduce the
\emph{downsizing} using the simulation based on the hierarchical
merging scenario and the $\Lambda$CDM model.

Interestingly, the BEGs in our sample also show some relationship
with the \emph{downsizing} model in their size -- redshift
relation, as shown in Fig.~9 and Fig.~10: the BEGs at lower
redshift are smaller than the BEGs at higher redshift, while the
size of the REGs does not vary significantly with redshift.
Because the BEGs have very similar morphological properties to
the REGs (elliptical shape and $R^{1/4}$ surface brightness
profile) and because there is no conspicuous discontinuity
between the REGs and the BEGs (as shown in Fig.~2 and Fig.~3), it
may be concluded that the BEGs will evolve into REGs. Evidence of
recent star-formation in nearby early-type galaxies also supports
this idea \citep{fer06,kav06}. On that assumption, the size --
redshift relation of BEGs seems to indicate that the mass of the
BEG-REG migration decreases as time goes, supporting the
`quenching mass decreasing with time' suggested by \citet{fab06}.
According to this scenario, BEGs may be the intermediate objects
from blue galaxies (late-type galaxies) to red galaxies
(early-type galaxies).

For small blue E/S0 galaxies, however, a new explanation was
suggested. \citet{fer05} regarded the faint blue E/S0 galaxies as
a different population from the bright blue E/S0 galaxies, on the
basis of the low surface brightness of the faint blue E/S0
galaxies. \citet{fer05} concluded that the faint blue E/S0
galaxies may be spirals with strong starbursts, whereas the bright
blue E/S0 galaxies will probably evolve into normal red E/S0
galaxies. However, if the small BEGs are really late-type
galaxies with very faint disks and spiral arms, the visible parts
of them are probably galactic bulges, whose color is typically
red due to old stellar populations \citep{zoc03} unlike the color
of the small BEGs. Another possible explanation is that they may
be the `progenitors' of late-type galaxies. According to the
spiral rebuilding' scenario suggested by \citet{ham05},
normal spiral galaxies may evolve and grow through the phases of
`merger' -- `compact galaxy' -- `disk growth', from $z\approx 1$
to $z\approx 0.4$. The small BEGs are possibly matched with
objects in the `compact galaxy' phase. However, this explanation
also has two weak points. First, many small BEGs are found at even
$z<0.4$ in our sample. In addition, the `spiral rebuilding'
scenario does not explain the size of the spiral bulges
decreasing with time, which is the opposite of the prediction of that
scenario.

A good possibility for the destiny of the small BEGs is that they
may evolve into dwarf elliptical galaxies. \citet{ilb06} also
predicted that `blue bulge-galaxies' may evolve into local dwarf
spheroidal galaxies. Interestingly, this
inference is consistent with the \emph{downsizing} scheme in the
formation of early-type galaxies. By analyzing the luminosity vs.
surface brightness relation of bright and dwarf elliptical
galaxies, \citet{gra03} showed that there is no intrinsic gap
between bright elliptical galaxies and dwarf elliptical galaxies,
and that they may be one continuous population just with different
masses, unlike the traditional belief that there is a dichotomy
between the two galaxy populations \citep{kor77,kor85}. In
addition, \citet{fer06} showed that there is no clear bimodality
between dwarf and regular elliptical galaxies in the surface
brightness profile and isophotal parameters, in their analysis of
100 early-type galaxies in the Virgo cluster using HST/ACS. In
this viewpoint, BEGs are probably the progenitors of normal and
dwarf elliptical galaxies, mostly originated from
mergers/interacting-galaxies or AGNs.

\section{CONCLUSION}

In this paper, we present a study of the properties of BEGs in the
GOODS HST/ACS fields. We selected 171 early-type galaxies visually
in the GOODS HST/ACS archival data with spectroscopic redshift.
We divided the early-type galaxies into 58 BEGs and 112 REGs
using their $(i-z)_{AB}$ color distribution at given redshift.
The BEGs have well-defined elliptical shapes and $R^{1/4}$
surface profiles similar to the REGs, with just a few exceptions.
However, the analysis of internal color distribution, X-ray
luminosity, and spectral line features show that almost a half of
the BEGs have evidence of tidal events and at least a quarter of
the BEGs probably contain an AGN in their centers. From the
analysis of the size and magnitude of the BEGs and the REGs, we
have found the evidence that the sizes of BEGs are decreasing as
redshift decreases, which is consistent with the \emph{downsizing}
scenario with hierarchical merging.

We conclude that the BEGs may be primarily descendants of past
merger/interacting-galaxies and secondarily the AGN-host
early-type galaxies, and that BEGs may evolve into normal or
dwarf elliptical galaxies. However, there are still several open
questions to answer. First, we could not find any BEGs at $z>1.2$ as
large as REGs. If larger REGs formed at the higher
redshift and if all REGs were BEGs at one time, many BEGs as
large as giant REGs are expected to exist at $z>1.2$. Therefore, the
question is whether all REGs evolved from BEGs or not. If not,
what is the difference between those two (or more) building
processes of early-type galaxies? The mechanism that makes the
central deep blue clumps in almost a half of the BEGs is another
question. It is hard to tell whether all \emph{blue clumpy} BEGs
contain AGNs or not, since the X-ray detection or matching with
optical catalog is limited. Deep observation in other
bands (e.g., mid-infrared spectroscopy using JWST \citep{rie05},
which may enables to identify the PAH emission of faint objects)
will be helpful to resolve which is dominant in the blue clumps
between AGN activity and star-forming activity. In addition, the
origin of the red inner rings in some BEGs is also a mystery to
be solved with further studies.

\acknowledgments

The authors are grateful to
Myungshin Im, Won-Kee Park, Narae Hwang, and Minjin Kim for
constructive suggestions, and to anonymous referee for very useful
comments.
This paper is in part supported by the
grant (R01-2004-000-10490-0) from the Basic Research Program of the 
Korea Science \& Engineering Foundation. 
This paper is based on observations with the NASA/ESA
\emph{Hubble space telescope}, obtained at the Space Telescope
Science Institute, which is operated by the Association of
Universities for Research in Astronomy, Inc. under NASA contract
NAS 5-26555. The HST/ACS observations are associated with
proposals 9425 and 9583. 

\appendix
\section{RING GALAXIES IN THE GOODS FIELDS}

We have found three ring galaxies in the GOODS fields, including
one BEG with an outer UV ring, GN11421 (GDS J123640.02+621207.7).
The other two ring galaxies, GN5542 (GDS J123618.94+620844.4) and
GN26250 (GDS J123729.85+621645.2) have very similar morphological
properties with GN11421. However, their rings can be identified
not only in the rest-frame UV band images but also in the
rest-frame optical band images. Therefore, GN5542 and GN26250 were
not included in our sample of early-type galaxies.

Fig.~17 shows the morphological variations along wavelength of the
three ring galaxies. The three ring galaxies look almost
spheroidal or elliptical in the $z$ band. The shorter the
wavelength is, however, the smaller their central spheroids are
and the brighter their outer rings appear. GN5542 and GN26250 show
more dramatic variations than GN11421: the central spheroids of
GN5542 and GN26250 almost disappear in their $B$ band images.
Assuming that the rings are circular and that they look elliptical
due to inclination, the diameters of the rings are estimated to be
5 kpc for GN5542 and GN11421, and 8 kpc for GN26250. These values
are relatively small compared to the sizes of nearby ring galaxies
\citep[5--52 kpc; ][]{the76}.

Ring galaxies in the nearby universe (e.g., the famous Cartwheel galaxy)
have been known for long. \citet{app96} explained that a ring
galaxy may be the result of a head-on collision between `a target
galaxy' and `an intruder'. According to their observational and
numerical studies, a ring wave that rotates and expands, occurs in
the target galaxy after the massive and compact intruder
penetrates the target galaxy. The $V$ band image of GN26250 is
surprisingly similar to that of the \emph{Cartwheel} galaxy that
is a representative collisional ring galaxy. Therefore these
distant ring galaxies were also probably formed by the same
process. To our knowledge, GN11421 ($z=1.015$) is the most (or at
least one of the most) distant ring galaxy ever found to date.
The most distant ring galaxy in the previous sample of
\citet{lav04} is at $z=0.996$.

\clearpage

\begin{deluxetable}{lcccccc}
\tablenum{1} \tablecolumns{7} \tablecaption{Basic Properties of
BEGs \label{tbl-1}} \tablewidth{0pt}

\tablehead{ID  & IAU ID & $z_{spec}$ & $B_{435}$ & $V_{606}$ &
$i_{775}$ & $z_{850}$ \\
(1) & (2) & (3) & (4) & (5) & (6) & (7) }

\startdata

GN 1430 &   GDS J123557.62+621024.7 &   3.068   &   23.812  &   22.714  &   22.572  &   22.622  \\
GN 3390 &   GDS J123603.62+621513.2 &   0.784   &   24.585  &   23.296  &   21.941  &   21.490  \\
GN 5302 &   GDS J123617.99+621635.3 &   0.679   &   21.973  &   21.152  &   20.460  &   20.183  \\
GN 7756 &   GDS J123627.32+621258.1 &   1.218   &   23.776  &   23.321  &   22.773  &   22.211  \\
GN 7904 &   GDS J123627.86+621124.8 &   0.517   &   22.983  &   22.304  &   21.469  &   21.160  \\
GN 8292 &   GDS J123629.44+621513.3 &   3.652   &   25.920  &   24.623  &   23.886  &   23.701  \\
GN 9126 &   GDS J123632.56+620800.2 &   1.994   &   24.110  &   24.249  &   23.747  &   23.563  \\
GN 10598    &   GDS J123637.32+620831.2 &   0.972   &   23.821  &   23.403  &   22.380  &   21.704  \\
GN 11103    &   GDS J123638.98+620912.1 &   0.342   &   23.740  &   23.170  &   22.870  &   22.781  \\
GN 11216    &   GDS J123639.30+621552.6 &   0.521   &   25.109  &   24.021  &   23.290  &   23.089  \\
GN 11421    &   GDS J123640.02+621207.7 &   1.015   &   24.729  &   23.614  &   22.429  &   21.694  \\
GN 12216    &   GDS J123642.49+621543.2 &   2.450   &   25.677  &   23.844  &   22.774  &   22.458  \\
GN 13334    &   GDS J123646.34+621405.0 &   0.961   &   23.873  &   22.657  &   21.571  &   20.865  \\
GN 14128    &   GDS J123649.06+621512.7 &   0.457   &   24.418  &   23.468  &   22.886  &   22.665  \\
GN 14229    &   GDS J123649.37+621311.6 &   0.476   &   23.594  &   22.774  &   22.207  &   22.007  \\
GN 14242    &   GDS J123649.43+621347.2 &   0.852   &   20.140  &   18.831  &   18.262  &   17.992  \\
GN 15003    &   GDS J123651.65+620954.9 &   0.136   &   22.434  &   21.777  &   21.406  &   21.334  \\
GN 15671    &   GDS J123653.52+622018.9 &   1.023   &   26.015  &   25.121  &   23.853  &   23.159  \\
GN 16895    &   GDS J123657.34+621026.4 &   0.846   &   25.135  &   24.143  &   22.955  &   22.479  \\
GN 17217    &   GDS J123658.30+621214.5 &   1.020   &   24.053  &   23.845  &   23.327  &   23.069  \\
GN 17606    &   GDS J123659.48+620815.0 &   0.116   &   22.034  &   21.365  &   21.027  &   20.927  \\
GN 19857    &   GDS J123706.96+621208.4 &   0.693   &   24.878  &   24.615  &   24.004  &   23.731  \\
GN 19996    &   GDS J123707.49+622148.1 &   1.451   &   22.971  &   22.803  &   22.382  &   22.026  \\
GN 21633    &   GDS J123713.22+621405.0 &   0.441   &   24.644  &   23.427  &   22.795  &   22.602  \\
GN 21983    &   GDS J123714.39+621221.5 &   1.084   &   25.353  &   24.500  &   23.608  &   22.895  \\
GN 24520    &   GDS J123723.19+621538.7 &   2.240   &   24.784  &   24.477  &   24.044  &   23.736  \\
GN 24667    &   GDS J123723.72+622113.3 &   3.524   &   25.174  &   23.957  &   23.634  &   23.621  \\
GN 25431    &   GDS J123726.64+621104.4 &   0.212   &   22.680  &   21.713  &   21.289  &   21.140  \\
GN 25594    &   GDS J123727.34+621319.2 &   0.513   &   22.973  &   22.513  &   21.958  &   22.107  \\
GN 28407    &   GDS J123739.49+621847.8 &   0.337   &   23.591  &   22.762  &   22.554  &   22.352  \\
GN 31118    &   GDS J123757.31+621627.8 &   2.922   &   23.685  &   22.471  &   21.982  &   21.872  \\
GS 467  &   GDS J033200.35$-$274319.7 &   1.037   &   22.705  &   22.348  &   22.214  &   22.033  \\
GS 895  &   GDS J033202.71$-$274310.8 &   0.493   &   20.425  &   19.000  &   18.438  &   18.164  \\
GS 1036 &   GDS J033203.29$-$274511.4 &   0.543   &   25.146  &   24.433  &   23.781  &   23.616  \\
GS 1222 &   GDS J033204.10$-$274424.0 &   0.117   &   24.173  &   23.763  &   23.604  &   23.649  \\
GS 3739 &   GDS J033210.91$-$274414.9 &   1.615   &   23.685  &   22.951  &   22.371  &   22.371  \\
GS 5632 &   GDS J033214.68$-$274337.1 &   0.976   &   25.121  &   24.364  &   23.478  &   23.017  \\
GS 6321 &   GDS J033215.98$-$274422.9 &   0.735   &   24.570  &   23.140  &   21.889  &   21.502  \\
GS 6587 &   GDS J033216.49$-$275019.9 &   0.407   &   24.859  &   23.553  &   22.899  &   22.729  \\
GS 6961 &   GDS J033217.14$-$274303.3 &   0.569   &   23.103  &   21.744  &   20.840  &   20.566  \\
GS 7517 &   GDS J033218.05$-$275000.8 &   1.474   &   24.924  &   24.356  &   23.693  &   23.465  \\
GS 8047 &   GDS J033219.00$-$274755.5 &   0.670   &   26.385  &   24.929  &   23.759  &   23.459  \\
GS 9479 &   GDS J033221.51$-$275359.7 &   0.966   &   25.810  &   24.502  &   23.775  &   23.348  \\
GS 10754    &   GDS J033223.66$-$275600.7 &   0.495   &   24.439  &   23.295  &   22.871  &   22.752  \\
GS 11796    &   GDS J033225.29$-$274224.2 &   0.613   &   24.595  &   23.145  &   22.435  &   22.239  \\
GS 12120    &   GDS J033225.77$-$274501.8 &   0.266   &   23.045  &   22.487  &   22.205  &   22.198  \\
GS 13482    &   GDS J033227.62$-$274144.9 &   0.665   &   24.050  &   22.803  &   21.613  &   21.302  \\
GS 17085    &   GDS J033232.96$-$274545.7 &   0.957   &   22.714  &   21.051  &   20.164  &   19.838  \\
GS 20048    &   GDS J033237.46$-$274000.1 &   0.666   &   23.524  &   22.902  &   22.155  &   21.856  \\
GS 20244    &   GDS J033237.76$-$275212.3 &   1.603   &   25.215  &   24.831  &   24.111  &   23.540  \\
GS 20457    &   GDS J033238.12$-$273944.8 &   0.837   &   20.934  &   20.526  &   20.630  &   20.432  \\
GS 21636    &   GDS J033240.04$-$274418.6 &   0.522   &   23.360  &   21.780  &   20.900  &   20.623  \\
GS 22653    &   GDS J033241.85$-$275202.5 &   3.610   &   24.350  &   22.885  &   22.403  &   22.430  \\
GS 23362    &   GDS J033243.24$-$274914.2 &   0.743   &   22.405  &   22.893  &   22.543  &   22.485  \\
GS 24938    &   GDS J033246.73$-$275352.9 &   0.214   &   24.975  &   24.088  &   23.665  &   23.563  \\
GS 27149    &   GDS J033252.39$-$275105.6 &   0.574   &   23.751  &   23.332  &   22.796  &   22.815  \\
GS 27611    &   GDS J033253.65$-$275319.0 &   0.734   &   25.033  &   24.571  &   23.894  &   23.697  \\
GS 28929    &   GDS J033258.71$-$275206.2 &   0.406   &   24.902  &   23.766  &   23.212  &   23.031  \\

\enddata
\tablecomments{ (1) Object MOSAIC ID number in the GOODS HST/ACS
official catalog ver1.1 (`GN' for GOODS-HDFN and `GS' for
GOODS-CDFS); (2) Official name of IAU; (3) Spectroscopic redshift;
(4) -- (7) AB AUTO magnitudes provided in the GOODS HST/ACS
official catalog ver1.1. The maximum photometric errors are
0.055, 0.027, 0.027, and 0.015 for $B,V,i$, and $z$,
respectively.}
\end{deluxetable}

\clearpage

\begin{deluxetable}{lccrccc}
\tablenum{2} \tablecolumns{7} \tablecaption{Additional Properties
of BEGs\label{tbl-2}} \tablewidth{0pt}
\tablehead{ID & $m-M$ & $M_z$ & $R_{hl,z}$ & $R_{hl,z}$ & $\Delta(i-z)_{SSP}$ & $\Delta(i-z)_{crit}$ \\
& & & (pixel) & (kpc) & & \\
(1) & (2) & (3) & (4) & (5) & (6) & (7) }

\startdata

GN 1430 &   47.09   &   $   -24.47  $   &   4.5     &   1.034  & 0.65 &   ---    \\
GN 3390 &   43.45   &   $   -21.96  $   &   8.0     &   1.793   & 0.16 &  0.05    \\
GN 5302 &   43.07   &   $   -22.89  $   &   20.3    &   4.297   & 0.20 &  0.09    \\
GN 7756 &   44.63   &   $   -22.42  $   &   7.5     &   1.875   & 0.30 &  ---    \\
GN 7904 &   42.35   &   $   -21.19  $   &   34.8    &   6.494   & 0.09 &  0.00    \\
GN 8292 &   47.54   &   $   -23.84  $   &   2.8     &   0.595   & 0.53 &  ---   \\
GN 9126 &   45.95   &   $   -22.39  $   &   5.5     &   1.383   & 1.14 &  ---    \\
GN 10598    &   44.03   &   $   -22.33  $   &   14.7    &   3.501   & 0.17 &  0.04    \\
GN 11103    &   41.29   &   $   -18.51  $   &   9.0     &   1.318   & 0.24 &  0.14    \\
GN 11216    &   42.37   &   $   -19.28  $   &   7.8     &   1.458   & 0.20 &  0.11    \\
GN 11421    &   44.14   &   $   -22.45  $   &   8.6     &   2.087   & 0.14 &  0.02    \\
GN 12216    &   46.50   &   $   -24.04  $   &   6.2     &   1.510   & 0.29 &  ---    \\
GN 13334    &   44.00   &   $   -23.14  $   &   14.8    &   3.516   & 0.14 &  0.01    \\
GN 14128    &   42.03   &   $   -19.37  $   &   6.8     &   1.187   & 0.16 &  0.07    \\
GN 14229    &   42.14   &   $   -20.13  $   &   6.9     &   1.230   & 0.19 &  0.10    \\
GN 14242    &   43.67   &   $   -25.68  $   &   43.5    &   10.021  & 0.48 &  0.36    \\
GN 15003    &   39.05   &   $   -17.72  $   &   12.3    &   0.892   & 0.24 &  0.08    \\
GN 15671    &   44.17   &   $   -21.01  $   &   5.1     &   1.239   & 0.19 &  0.07    \\
GN 16895    &   43.65   &   $   -21.17  $   &   6.3     &   1.436   & 0.23 &  0.11    \\
GN 17217    &   44.16   &   $   -21.09  $   &   6.8     &   1.649   & 0.62 &  0.50    \\
GN 17606    &   38.68   &   $   -17.75  $   &   14.1    &   0.899   & 0.20 &  0.03    \\
GN 19857    &   43.12   &   $   -19.39  $   &   4.0     &   0.858   & 0.20 &  0.09    \\
GN 19996    &   45.10   &   $   -23.07  $   &   6.5     &   1.650   & 0.45 &  ---    \\
GN 21633    &   41.94   &   $   -19.34  $   &   9.1     &   1.566   & 0.18 &  0.09    \\
GN 21983    &   44.32   &   $   -21.43  $   &   5.9     &   1.440   & 0.22 &  0.11    \\
GN 24520    &   46.26   &   $   -22.52  $   &   2.7     &   0.671   & 0.61 &  ---    \\
GN 24667    &   47.45   &   $   -23.83  $   &   3.0     &   0.662   & 0.61 &  ---   \\
GN 25431    &   40.11   &   $   -18.97  $   &   20.0    &   2.083   & 0.15 &  0.02    \\
GN 25594    &   42.33   &   $   -20.22  $   &   6.7     &   1.249   & 0.55 &  0.46    \\
GN 28407    &   41.26   &   $   -18.91  $   &   11.5    &   1.668   & 0.12 &  0.02    \\
GN 31118    &   46.96   &   $   -25.09  $   &   4.3     &   0.991   & 0.40 &  ---    \\
GS 467  &   44.20   &   $   -22.17  $   &   5.0     &   1.213   & 0.72 &  0.60    \\
GS 895  &   42.23   &   $   -24.07  $   &   51.0    &   9.295   & 0.12 &  0.03    \\
GS 1036 &   42.48   &   $   -18.86  $   &   5.3     &   1.010   & 0.24 &  0.15    \\
GS 1222 &   38.70   &   $   -15.05  $   &   5.3     &   0.336   & 0.34 &  0.17    \\
GS 3739 &   45.39   &   $   -23.02  $   &   4.8     &   1.209   & 0.96 &  ---    \\
GS 5632 &   44.04   &   $   -21.02  $   &   6.4     &   1.531   & 0.40 &  0.27    \\
GS 6321 &   43.28   &   $   -21.78  $   &   13.8    &   3.010   & 0.14 &  0.03    \\
GS 6587 &   41.74   &   $   -19.01  $   &   3.3     &   0.531   & 0.19 &  0.10    \\
GS 6961 &   42.60   &   $   -22.03  $   &   11.0    &   2.157   & 0.14 &  0.04    \\
GS 7517 &   45.15   &   $   -21.69  $   &   2.6     &   0.661   & 0.58 &  ---    \\
GS 8047 &   43.03   &   $   -19.57  $   &   5.0     &   1.051   & 0.16 &  0.05    \\
GS 9479 &   44.01   &   $   -20.66  $   &   4.0     &   0.955   & 0.42 &  0.29    \\
GS 10754    &   42.24   &   $   -19.49  $   &   3.8     &   0.685   & 0.27 &  0.18    \\
GS 11796    &   42.80   &   $   -20.56  $   &   4.0     &   0.812   & 0.23 &  0.13    \\
GS 12120    &   40.67   &   $   -18.47  $   &   4.0     &   0.493   & 0.30 &  0.19    \\
GS 13482    &   43.02   &   $   -21.72  $   &   8.0     &   1.683   & 0.15 &  0.05    \\
GS 17085    &   43.99   &   $   -24.15  $   &   16.5    &   3.928   & 0.52 &  0.39    \\
GS 20048    &   43.02   &   $   -21.16  $   &   8.6     &   1.800   & 0.16 &  0.06    \\
GS 20244    &   45.37   &   $   -21.83  $   &   4.9     &   1.240   & 0.39 &  ---    \\
GS 20457    &   43.63   &   $   -23.20  $   &   9.3     &   2.119   & 0.51 &  0.39    \\
GS 21636    &   42.38   &   $   -21.76  $   &   13.5    &   2.535   & 0.12 &  0.03    \\
GS 22653    &   47.51   &   $   -25.08  $   &   4.0     &   0.870   & 0.66 &  ---   \\
GS 23362    &   43.31   &   $   -20.83  $   &   4.8     &   1.044   & 0.46 &  0.35    \\
GS 24938    &   40.13   &   $   -16.57  $   &   4.8     &   0.506   & 0.20 &  0.07    \\
GS 27149    &   42.63   &   $   -19.82  $   &   5.5     &   1.075   & 0.43 &   0.33    \\
GS 27611    &   43.28   &   $   -19.58  $   &   6.3     &   1.379   & 0.32 &  0.21    \\
GS 28929    &   41.73   &   $   -18.70  $   &   6.3     &   1.030   & 0.18 &  0.09    \\

\enddata
\tablecomments{ (1) Object ID; (2) Distance modulus; (3) Absolute
magnitude in the z band (observer-frame, NOT rest-frame); (4) --
(5) Half-light radius in the z band (in pixel and in kpc); (6)
$(i-z)$ color difference from the expected value in the SSP model
\citep{bru03} with [Fe/H]$=-0.4$ and $z_{F}=5$; (7) $(i-z)$ color
difference from the selection criteria for BEGs (at $z<1.2$). }
\end{deluxetable}

\clearpage

\begin{deluxetable}{lllll}

 \tablenum{3} \tablecolumns{5} \tablecaption{Structural, X-ray, and Spectral Features of
 BEGs}
 \label{tbl-3} \tablewidth{0pt}  \tablehead{ID & Central color distribution & Outer color distribution & log $L_{X-ray}$ &
 Spectrum \\
 & & & (erg s$^{-1}$) & \\
 (1) & (2) & (3) & (4) & (5) }

\startdata

GN 1430 &   symmetric deep BC  &   tidally distorted   &  43.98   &   --- \\
GN 3390 &   little variation &   tidally distorted   &   --- &   --- \\
GN 5302 &   symmetric deep BC  &   very nearby objects    &    43.35   &   --- \\
GN 7756 &   asymmetric? shallow BC    &   very nearby objects    &   42.01   &   --- \\
GN 7904 &   little variation &   very nearby objects    &   --- &   --- \\
GN 8292 &   asymmetric deep BC   &   --- &    44.35   &   --- \\
GN 9126 &   asymmetric? shallow BC    &   tidally distorted   &  42.91   &   --- \\
GN 10598    &   off-center BC   &   tidally distorted   &   41.67   &   --- \\
GN 11103    &   symmetric shallow BC    &   --- &   --- &    --- \\
GN 11216    &   little variation &   --- &   --- &   --- \\
GN 11421    &   little variation &   blue outer ring &   --- &   --- \\
GN 12216    &   little variation &   --- &   --- &   --- \\
GN 13334    &   symmetric deep BC  &   --- &     43.57   &   --- \\
GN 14128    &   symmetric shallow BC &   --- &   --- &   --- \\
GN 14229    &   symmetric shallow BC    &   --- &    --- &   --- \\
GN 14242    &   flat    &   very nearby objects    &   41.53   &   --- \\
GN 15003    &   asymmetric? shallow BC   &   --- &  --- &   --- \\
GN 15671    &   symmetric shallow BC & tidally distorted &  --- &   --- \\
GN 16895    &   off-center BC &   ---  &   --- &   --- \\
GN 17217    &   off-center BC   &   --- &   --- &   --- \\
GN 17606    &   little variation &   --- &  --- &   --- \\
GN 19857    &   asymmetric shallow BC   &   --- &   --- &   --- \\
GN 19996    &   symmetric deep BC  &   red inner ring  &  44.42   &   --- \\
GN 21633    &   asymmetric? shallow BC    &   --- &   --- &  --- \\
GN 21983    &   asymmetric shallow BC    &   --- & --- &   --- \\
GN 24520    &   asymmetric shallow BC   &   --- &    43.53   &   --- \\
GN 24667    &   asymmetric? deep BC   &   --- &   44.12   &   --- \\
GN 25431    &   flat    &   --- &    --- &   --- \\
GN 25594    &   off-center BC   &   --- &   --- &   --- \\
GN 28407    &   multiple BC   &   --- &    --- &   --- \\
GN 31118    &   asymmetric? deep BC  &   red inner ring  &   44.58   &   --- \\
GS 467  &   symmetric deep BC  &   red inner ring  &   43.69   &   AGN \\
GS 895  &   flat    &   --- &   --- &   absorption   \\
GS 1036 &   asymmetric? shallow BC    &   --- &   --- &   emission    \\
GS 1222 &   asymmetric? shallow BC    &   --- &   --- &   absorption   \\
GS 3739 &   asymmetric? deep BC   &   red inner ring  &  44.40   &   AGN \\
GS 5632 &   symmetric shallow BC    &   --- &  --- &   composite   \\
GS 6321 &   little variation &   --- &   --- &  absorption   \\
GS 6587 &   asymmetric deep BC   &   red inner ring  & --- &   absorption?   \\
GS 6961 &   symmetric shallow BC    &   --- &  43.14   &   AGN(+absorption?) \\
GS 7517 &   off-center BC   &   --- & --- &   absorption   \\
GS 8047 &   off-center BC   &   --- &  42.18   &   absorption   \\
GS 9479 &   asymmetric shallow BC    &   --- &  --- &   emission?    \\
GS 10754    &   symmetric deep BC  &   red inner ring  &    --- &   absorption   \\
GS 11796    &   asymmetric deep BC  &   red inner ring  &  --- &   composite   \\
GS 12120    &   asymmetric? deep BC  &   red inner ring  &    --- &   composite    \\
GS 13482    &   symmetric shallow BC    &   --- &  42.35   &   AGN    \\
GS 17085    &   flat    &   very nearby objects    & --- &   absorption   \\
GS 20048    &   symmetric deep BC  &   tidally distorted   &  42.93   &   AGN   \\
GS 20244    &   asymmetric? shallow BC    &   --- &  44.38   &   AGN \\
GS 20457    &   symmetric deep BC  &   red inner ring  &  --- &   AGN \\
GS 21636    &   symmetric shallow BC    &   --- & --- &   --- \\
GS 22653    &   symmetric deep BC   &   red inner ring  & 44.65   &   AGN    \\
GS 23362    &   symmetric deep BC   &   red inner ring  & 43.34   &   composite   \\
GS 24938    &   asymmetric shallow BC    &   --- &  --- &   emission    \\
GS 27149    &   asymmetric shallow BC   &   --- &  --- &   emission    \\
GS 27611    &   little variation &   --- &   --- &emission    \\
GS 28929    &   symmetric shallow BC    &   --- & --- &   composite   \\

\enddata

\tablecomments{ (1) Object ID; (2) Simplified internal color
distribution feature, especially about the existence of blue
clumps (BCs); (3) Simplified color distribution feature,
especially about tidal structures and rings; (4) X-ray luminosity
in the rest-frame 0.5keV -- 10keV band, calculated with a typical
AGN SED; (5) Spectral line feature (CDFS only). }

\end{deluxetable}


\begin{figure}
\epsscale{.80} \plotone{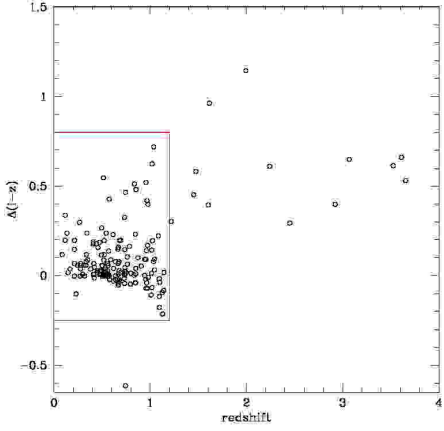} \caption{ The $(i-z)_{\rm AB}$
color differences ($\Delta (i-z) \equiv (i-z)_{\rm model} -
(i-z)_{\rm obs}$) vs. redshift, of 171 early-type galaxies. To
derive $(i-z)_{\rm model}$, we used the SSP model \citep{bru03}
with $z_{\rm F}=5$ and [Fe/H]$=-0.4$. The box represents the
domain used in Fig. 2. \label{fig1}}
\end{figure}


\begin{figure}
\epsscale{.80} \plotone{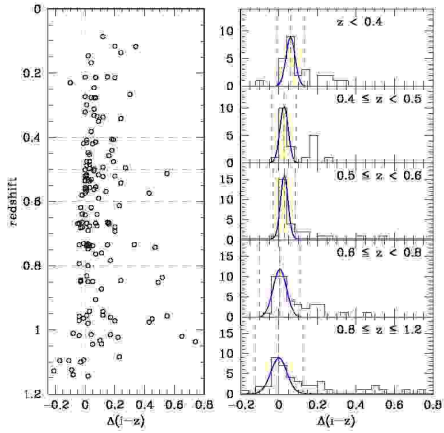} \caption{\emph{Left panel}: The
redshift vs. $\Delta (i-z)$ for $0 \le z \le 1.2$. We divided the
redshift range into five bins so that each bin contains a similar
number of galaxies (separated by four horizontal dotted lines).
\emph{Right panel}: Color histograms for five redshift bins.
Curved lines represent gaussian fits. Three vertical dashed-lines
represent peak$- 3 \sigma$, gaussian peak, and peak$+ 3 \sigma$
from left to right. \label{fig2}}
\end{figure}


\begin{figure}
\epsscale{.80} \plotone{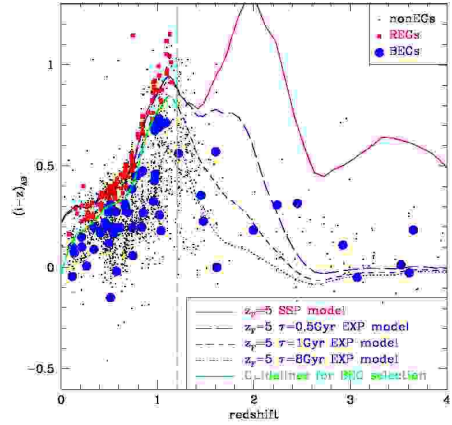} \caption{The $(i-z)$ color vs.
redshift of 113 REGs (filled squares), 58 BEGs (filled circles),
and 1,778 non-early-type galaxies (dots). Several evolutionary
models are overlaid: SSP model with ${\rm z}_{\rm F}=5$ (solid
line), and exponentially-decreasing star-formation rate models
(`EXP models') with ${\rm z}_{\rm F}=5$ and $\tau=$ 0.5 Gyr
(long-dashed line), 1 Gyr (short-dashed line), and 8 Gyr (dotted
line). We adopted [Fe/H] $=-0.4$ in all models. The dot-dashed
lines are the guidelines for REG-BEG separation.\label{fig3}}
\end{figure}


\begin{figure}
\epsscale{.80} \plotone{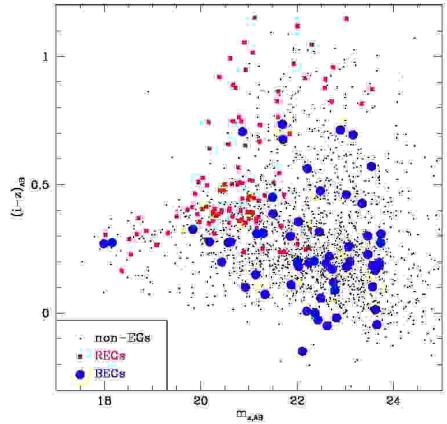} \caption{ The $(i-z)_{\rm AB}$
vs. $z_{\rm AB}$ of the BEGs (filled circles) in comparison with
the REGs (filled squares) and non-early-type galaxies (dots).
\label{fig4}}
\end{figure}


\begin{figure}
\epsscale{.80} \plotone{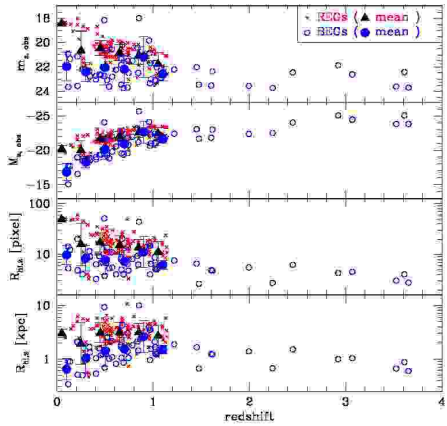} \caption{ The magnitude and size
distributions of the REGs (crosses) and the BEGs (open circles)
as functions of redshift: apparent $z$-band magnitude, absolute
$z$-band magnitude, half-light radius in pixel, and half-light
radius in kpc from upper to lower. The mean values are also
presented (filled triangles for the REGs and filled circles for
the BEGs). The magnitudes are {\emph not} k-corrected, because
the SED of the BEGs is not well known. \label{fig5}}
\end{figure}


\begin{figure}
\epsscale{.80} \plotone{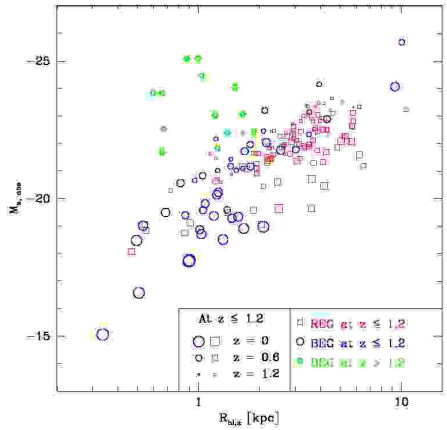} \caption{ The absolute magnitude
vs. the half-light radius [kpc] of the REGs (open squares) and
the BEGs (open circles and open stars). For the early-type
galaxies at $z\le 1.2$, the symbol sizes decrease as redshift
increases, and the BEGs at $z>1.2$ are denoted as open stars of
constant symbol size. \label{fig6}}
\end{figure}


\begin{figure}
\epsscale{.80} \plotone{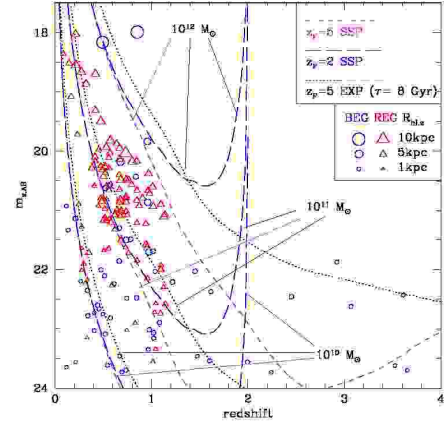} \caption{ The magnitude vs.
redshift of the REGs (open triangles) and the BEGs (open circles)
with several galaxy evolutionary tracks from the model
\citep{bru03} with [Fe/H] $= -0.4$. Short-dashed lines,
long-dashed lines and dotted lines represent, respectively, the
SSP models of single burst at $z_{F}=5$, single burst at
$z_{F}=2$ and exponential star-formation with $\tau=8$ Gyr. For
each model, three evolutionary lines for the masses of $10^{10}$,
$10^{11}$, $10^{12}$M$_{\odot}$ are drawn. Symbol sizes of the
BEGs and the REGs are proportional to their half-light radii.
\label{fig7}}
\end{figure}

\begin{figure}
\epsscale{.80} \plotone{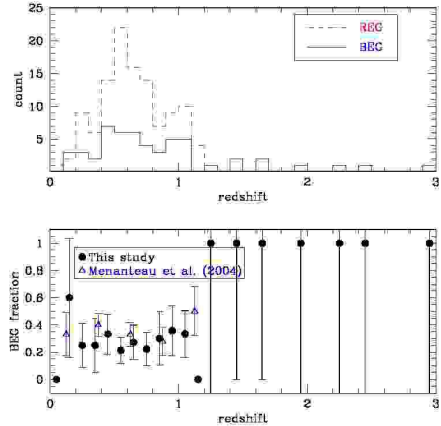} \caption{ \emph{Upper panel}: The
number counts of the REGs (dashed line) and the BEGs (solid
line). \emph{Lower panel}: The variation of BEG fraction among
entire early-type galaxies as a function of redshift in this study
(filled circle) and in \citet{men04} (open triangle).
\label{fig8}}
\end{figure}


\begin{figure}
\epsscale{.80} \plotone{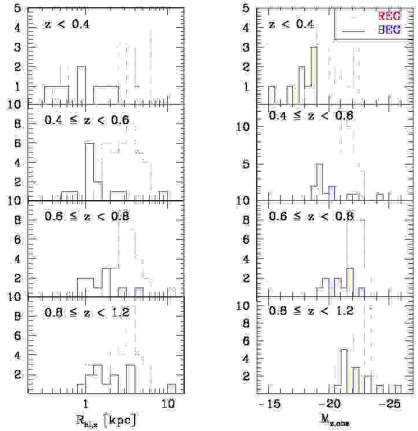} \caption{ The half-light radius
distribution (left panel) and the absolute magnitude distribution
(right panel) of the REGs (dotted line) and the BEGs (solid line)
for four redshift bins. \label{fig9}}
\end{figure}


\begin{figure}
\epsscale{.80} \plotone{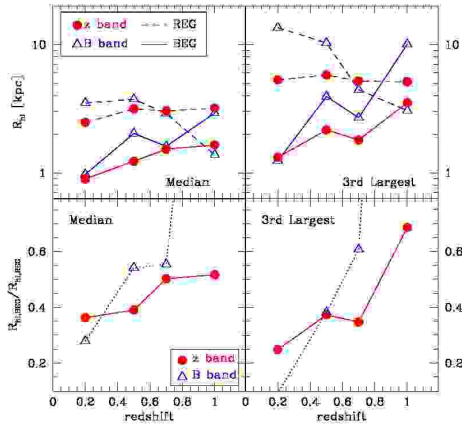} \caption{ \emph{Upper-left
panel}: The median sizes of the REGs and the BEGs as functions of
redshift. \emph{Lower-left panel}: The ratio of BEG median size
to REG median size as a function of redshift. \emph{Upper-right
panel}: The 3rd largest galaxy sizes of the REGs and the BEGs as
functions of redshift. \emph{Lower right panel}: The 3rd largest
galaxy size ratio of BEGs to REGs as a function of redshift.
Filled circles represent $z$ band and open triangles represent
$B$ band in all panels. In the two upper panels, solid lines
represent the values for the BEGs and dashed lines for the REGs.
\label{fig10}}
\end{figure}


\begin{figure}
\epsscale{1.00} \plotone{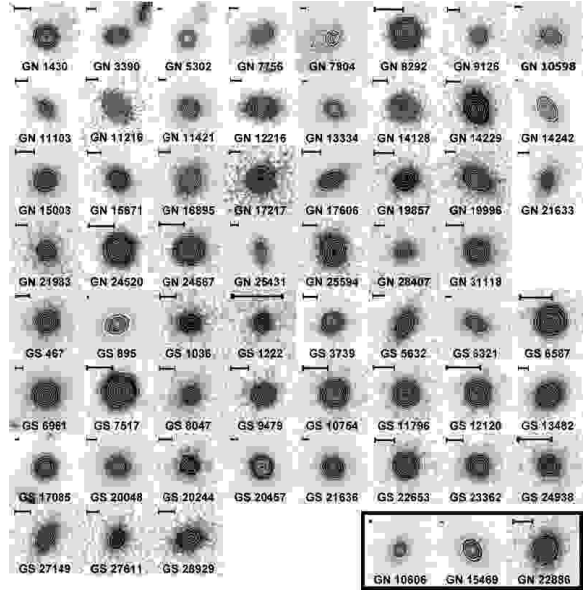} \caption{ The gray-scale maps
and isophotal contours of $z$-band images of the 58 BEGs and
three typical REGs with different apparent magnitudes (GN10606,
GN15469, and GN22886 at the lower-right corner; $m_{z} = 20.85$,
18.87, and 22.01 respectively). The size of each image was
adjusted for the best view of each galaxy based on its half-light
radius, with a scale-bar indicating a physical length, 2 kpc.
\label{fig11}}
\end{figure}


\begin{figure}
\epsscale{1.00} \plotone{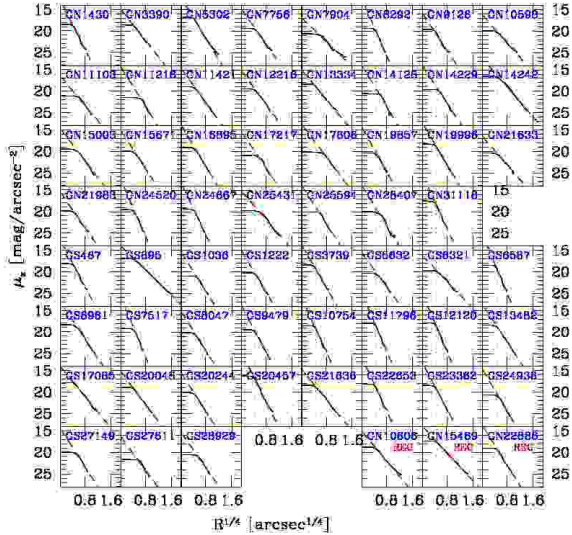} \caption{ The surface brightness
profiles of the 58 BEGs and three REGs for comparison in the $z$
band. The outer part of each profile is overlaid with a $R^{1/4}$
fitting line (dashed line), and the average PSF HWHM (half width
half maximum) is denoted as a vertical dotted line. \label{fig12}}
\end{figure}


\begin{figure}
\epsscale{1.00} \plotone{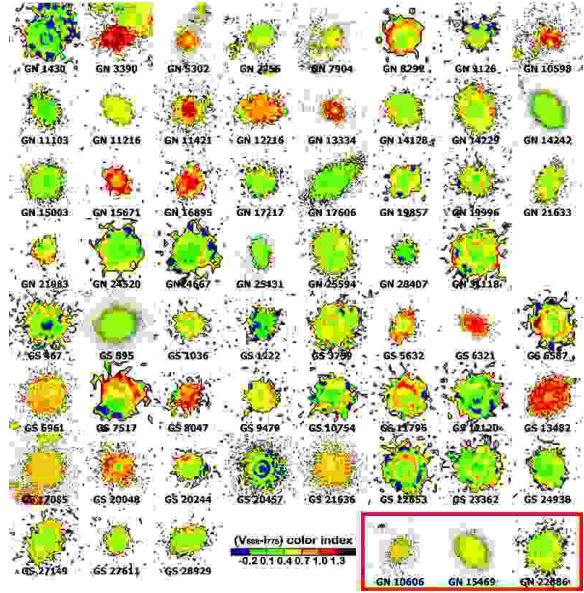} \caption{ The pseudo-color maps
representing the $(V-i)$ color of the BEGs, with the same sizes as
the portrait images in Fig.~11. In the lower-right box, the
pseudo-color images of three typical REGs are shown for
comparison. The bar in the lower-center shows the $(V-i)$ range
that each color tone represents. \label{fig13}}
\end{figure}


\begin{figure}
\epsscale{1.00} \plotone{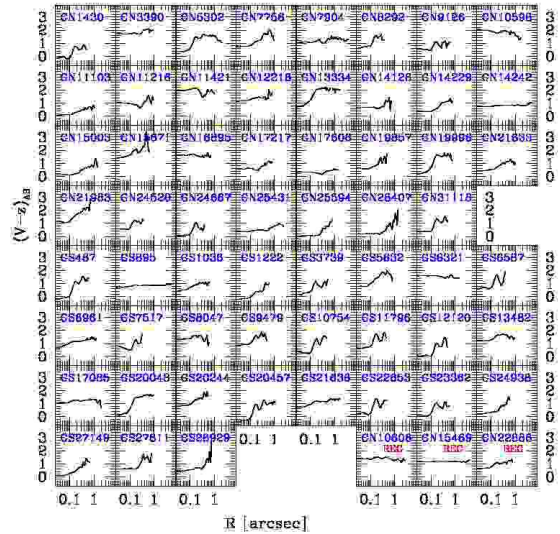} \caption{ The $(V-z)$ color
profiles of the 58 BEGs and three REGs. The PSF HWHM is $0.05$
arcsec. \label{fig14}}
\end{figure}


\begin{figure}
\epsscale{.80} \plotone{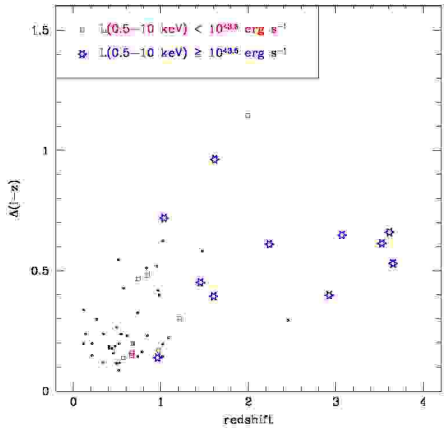} \caption{ The X-ray luminosity
features on the $\Delta (i-z)$ vs. redshift diagram. Open squares
and open stars represent, respectively, high X-ray luminosity
BEGs (L$_{0.5-10 {\rm keV}} \ge 10^{43.5}$ erg ${\rm s}^{-1}$)
and low X-ray luminosity BEGs (L$_{0.5-10 {\rm keV}} < 10^{43.5}$
erg ${\rm s}^{-1}$). The BEGs that were not detected in the X-ray
are also plotted (small open circles). \label{fig15}}
\end{figure}


\begin{figure}
\epsscale{.80} \plotone{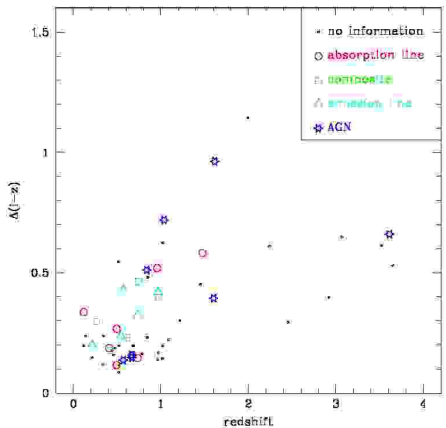} \caption{ The spectral line
features on the $\Delta (i-z)$ vs. redshift diagram. We
classified the BEGs according to their spectral line features into
four categories: absorption line galaxies (open large circles),
emission line galaxies (open triangles), composite line galaxies
(open squares), and AGNs (open stars). The BEGs without spectral
line information (small open circles) are also plotted.
\label{fig16}}
\end{figure}


\begin{figure}
\epsscale{.80} \plotone{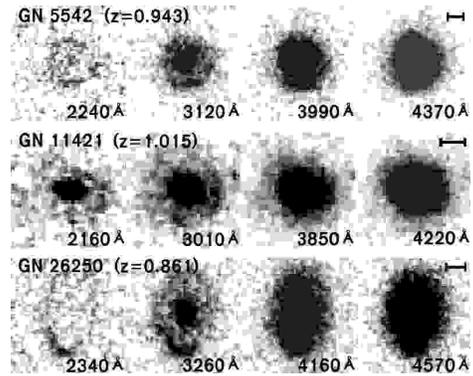} \caption{ Grayscale maps of four
band images of the ring galaxies. The lower-right number in each
image represents the rest-frame wavelength in the $B$, $V$, $i$,
and $z$ bands, respectively. The scale-bar in the upper-right
corner for each galaxy represents the length of 2 kpc.
\label{fig17}}
\end{figure}

\end{document}